\documentclass[a4paper,twocolumn,11pt,accepted=2025-02-28]{quantumarticle}
\pdfoutput=1
\usepackage[utf8]{inputenc}
\usepackage[english]{babel}
\usepackage{longtable}
\usepackage{booktabs}
\usepackage[T1]{fontenc}
\usepackage{amsmath}
\usepackage{amssymb}
\usepackage{hyperref}
\usepackage{braket}
\usepackage{tikz}
\usepackage{lipsum}
\usepackage[numbers]{natbib}
\usepackage{comment}

\newcommand{\sign}{\text{sign}}

\newtheorem{theorem}{Theorem}

\newcommand{\liubov}[1]{\textcolor{orange}{[liubov] #1}}

\begin{document}

\title{Quantum memory assisted observable estimation}

\author{Liubov Markovich}
\affiliation{Instituut-Lorentz, Universiteit Leiden, P.O. Box 9506, 2300 RA Leiden, The Netherlands}
\orcid{0000-0002-1360-7069}
\email{markovich@mail.lorentz.leidenuniv.nl}

\author{Attaallah Almasi}
\orcid{0000-0002-6425-6852}
\affiliation{QuTech and Kavli Institute of Nanoscience, Delft University of Technology, 2628 CJ, Delft, The Netherlands}

\author{Sina Zeytino\u{g}lu}
\affiliation{Physics and Informatics Laboratory, NTT Research, Inc.,
Sunnyvale, California, 94085, The USA}
\affiliation{Institute for Theoretical Physics, TU Wien, Wiedner Hauptstraße 8-10/136, A-1040
Vienna, Austria}

\orcid{0000-0002-5529-5681}

\author{Johannes Borregaard}

\affiliation{Department of Physics, Harvard University, Cambridge, MA 02138, USA}
\orcid{0000-0003-2544-4073}

\maketitle

\begin{abstract}
The estimation of many-qubit observables is an essential task of quantum information processing. The generally applicable approach is to decompose the observables into weighted sums of multi-qubit Pauli strings, i.e., tensor products of single-qubit Pauli matrices, which can readily be measured with low-depth Clifford circuits. The accumulation of shot noise in this approach, however, severely limits the achievable variance  for a finite number of measurements. We introduce a novel method, dubbed \textit{coherent Pauli summation} (CPS), that circumvents this limitation by exploiting access to a single-qubit quantum memory in which measurement information can be stored and accumulated. 
CPS offers a reduction in the required number of measurements for a given variance  that scales linearly with the number of Pauli strings in the decomposed observable. Our work demonstrates how a single long-coherence qubit memory can assist the operation of many-qubit quantum devices in a cardinal task.   
\end{abstract}
\section{Introduction}
Quantum devices with on the order of hundreds of qubits have been realized with superconducting hardware~\cite{jurcevic2021,kjaergaard2020superconducting}, neutral atoms~\cite{henriet2020,PhysRevLett.123.170503}, and trapped ions~\cite{brown2016,schafer2018,risinger2021characterization}. These advancements stimulated
interest in simulating many-body systems such as the electronic structure of molecules~\cite{Babbush2018}, 
 studying non-equilibrium quantum statistical mechanics~\cite{Yang2020}, and performing combinatorial optimization~\cite{Farhi2022quantumapproximate}
 on such devices. A cardinal task for many of these applications is to estimate the expectation values of many-qubit observables, such as the energy of the system. The direct estimation of such observables can be highly non-trivial for, e.g., fermionic observables simulated on qubit systems~\cite{RevModPhys.92.015003} 
 and poses a significant challenge due to large measurement circuit depths and overall sampling complexity, i.e., the total number of measurements for a required estimation variance.   

One approach for observable estimation with minimal sampling complexity is the quantum phase estimation (QPE) algorithm~\cite{Kitaev1995,Kimmel2015,Berg2020,mohammadbagherpoor2019,Wiebe2016,O_Brien2019}. Its implementation, however, requires qubit systems with low noise and long coherence times for high precision estimation since the measurement circuit depth is inversely proportional to the square root of the achievable estimation variance. In addition, the observable of interest has to be encoded as a unitary transformation which is, in general, a non-trivial requirement.
\par As an alternative, the quantum energy (expectation) estimation (QEE) approach of decomposing the observable into a weighted sum of $N$ multi-qubit Pauli strings is commonly used in the variational quantum eigensolver~\cite{Peruzzo2014}. While QEE minimizes the measurement circuit depth by requiring only a single layer of single-qubit rotations, it also suffers from increased sample complexity due to the accumulation of shot noise in the estimation procedure. Specifically, the expectation values of Pauli strings are estimated independently, and the observable is then calculated as a linear combination of these. Consequently, to estimate an observable comprised of $N$ Pauli stings to a variance  $\eta$, each Pauli string should be estimated to a variance  $O(\eta/N)$ resulting in an overall sample complexity scaling as $O(N^2)$.  
This accumulation of noise poses a ``shot noise bottleneck" since the amount of measurements will ultimately be limited
by the available runtime of the device before, e.g., re-calibration of the device is needed. The measurement process itself is also often one of the most time consuming operations in current quantum devices~\cite{jurcevic2021,brown2016,schafer2018}. 
\begin{center} 
    \begin{figure} [t]
        \includegraphics[width=8.3cm]{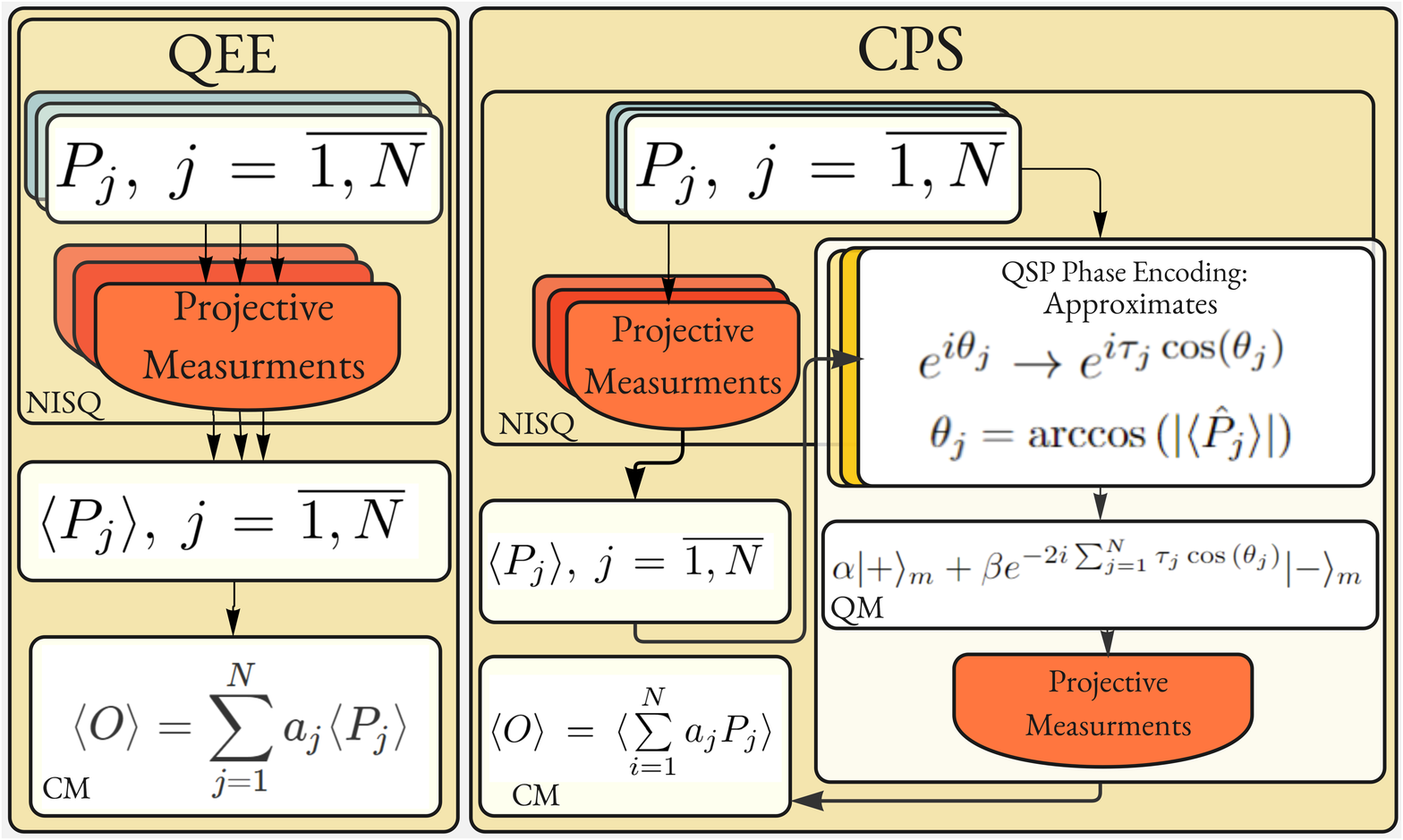}
         \caption{ The comparison of the standard QEE method and the proposed CPS method. On the left, the QEE, where the expectation value of each Pauli string ($\langle \hat{P}_j\rangle$, $j=1,\dots, N$) is estimated by a series of projective measurements. Finally, an estimate of the observable $\langle \hat{O}\rangle$ is obtained by a weighted summation of all $\langle \hat{P}_j\rangle$. On the right, the CPS, where $\langle \hat{P}_j\rangle$ is encoded in the single qubit quantum memory (QM)  such that a direct encoding of $\langle \hat{O}\rangle$ is obtained. To ensure the right summation of the Pauli strings in the phase of the memory qubit, quantum signal processing (QSP) is used for efficient approximation of the required function. In addition, a small amount of projective measurements of each Pauli string is also being performed for Pauli strings sign estimation ($\tau_j$) purposes. After the encoding process is done, a final projective measurement on the quantum memory qubit is performed. The procedure is then iterated to obtain an estimate of $\langle \hat{O}\rangle$ to the desired variance.
         \label{fig_overall}
        }
    \end{figure}  
\end{center}

To tackle the shot noise bottleneck, recent works have considered intermediate approaches between QPE and QEE to obtain better variance  in the estimation of the individual Pauli strings~\cite{Wang2019} or methods for grouping Pauli strings in commuting sets to reduce the sample complexity~\cite{Hamamura2020,crawford2021efficient}. While both approaches have the potential to reduce the overall sample complexity, neither improves the fundamental scaling of the noise accumulation with the number of Pauli strings in the observable decomposition.

Here, we propose a novel approach dubbed \textit{coherent Pauli summation} (CPS) that overcomes the shot-noise bottleneck through the use of a single-qubit quantum memory (QM) and Quantum Signal Processing (QSP) techniques \cite{PhysRevX.6.041067,gilyen2019quantum}. Our method allows for a direct measurement of the multi-qubit observable by estimating the phase of the memory qubit at the end of the protocol. Hence, the accumulation of shot noise, originating from the summation of individually estimated mean values of Pauli strings in the QEE approach, is prevented.  Importantly, the properties of the QSP-based encoding allow this task to be completed with processing qubits whose coherence time scales only logarithmically with the required estimation variance, in contrast to the linear scaling of the QPE algorithm. 

The performance of the QPE algorithm for observable estimation can significantly vary depending on the method.  For example, in Refs.~\cite{dutkiewicz2021,Higgins2009,Kimmel2015,Wiebe2016},
Heisenberg limited scaling is achieved assuming the observable can be encoded with a unitary evolution. This assumption is difficult to satisfy in practice and limits the applicability of the result especially for current stage of technological development. Other types of QPE algorithms relax this restrictive assumption but result in worse performance~\cite{O_Brien2019,Wang2019}. 
\par The critical point of the CPS method is to employ QSP techniques to encode the mean value of Pauli strings in the phase of a single qubit's quantum memory, enabling us to artificially rescale the phase between $0$ and $2\pi$ without additional cost~\cite{PhysRevX.6.041067}.
Furthermore, this allows us to obtain Heisenberg scaling of the estimation variance with circuit depths on the many-qubit device that increase only logarithmically with the inverse of the achievable estimation variance. This is because the many-qubit device can be readout and reset after each encoding into the memory qubit. The total circuit depth of the memory qubit is inversely proportional to the square root of the achievable estimation variance as in standard QPE. From this perspective, one can think of CPS as a new variation of QPE, using QSP and the single qubit QM to achieve the Heisenberg scaling without paying a high price of a unitary preparation or long circuits depths on the many-qubit device. Thus, CPS outlines how a single error corrected qubit memory can advance the performance of larger scale quantum devices. The lack of shot noise accumulation results in a gain in the variance of the estimate of $O(1/\sqrt{N})$ compared to performing QPE of individual Pauli strings which are summed classically~\cite{Wang2019} and the QPE scaling results in a gain of $O(1/T)$ where $T$ is the amount of state preparations compared to QEE.

\section{CPS method}
In order to set the stage of our algorithm, we first review the basic steps of the standard QEE approach (see also Fig.~\ref{fig_overall}). The first step of the QEE approach for estimating the expectation value of an observable $\hat{O}$ for a given quantum state $|\Psi\rangle$, is to decompose it into a weighted sum of Pauli strings
\begin{equation} \label{eq:0}
    \hat{O}=\sum\limits_{j=1}^Na_{j}\hat{P}_{j},
\end{equation}
where $a_j$ are the (real) decomposition coefficients and $\hat{P}_{j}$ are the Pauli strings composed as tensor products of single qubit Pauli matrices and the identity. This decomposition is always possible since a collection of $d^2$ Pauli strings forms a complete operator basis for a $d$ dimensional Hilbert space. However, the number of Pauli strings, $N$, in the decomposition can be very large for a general multi-qubit observable. For example, when mapping fermionic systems onto qubit quantum devices, local fermionic observables can map to multi-qubit observables spanning the device~\cite{PhysRevB.104.035118,nys2022quantum,Verstraete_2005}.

In general, the state $|\Psi\rangle$ is repeatedly prepared, and the Pauli strings are measured sequentially via projective measurements to obtain estimates of every Pauli string: $\langle \hat{P}_j\rangle\equiv\langle \Psi|\hat{P}_j|\Psi\rangle$~\cite{Peruzzo2014,McClean2014}. The mean value of the observable $ \hat{O}$ can be calculated classically after estimating all $\langle \hat{P}_j\rangle$, $j=[1, N]$. This approach, however, suffers from the accumulation of shot noise from the individually estimated mean values of the Pauli strings, as described above. If each estimate $\langle \hat{P}_j\rangle$ is estimated with a variance  $\sigma^2(\langle\hat{P}_j\rangle)= \eta_p$,  the variance of the final estimate is  $\sigma^2(\hat{\langle O\rangle})\approx N\eta_p$, assuming roughly equal weights of the $\hat{P}_j$'s in \eqref{eq:0}. 

We now outline the CPS method that circumvents this accumulation of shot noise (see Fig.~\ref{fig_overall}). Let $|\Psi_0\rangle \equiv \hat{V}|\pmb{0}\rangle$ be a quantum state of the  multi-qubit device, where $\hat{V}$ is an invertible preparation circuit. The state $|\pmb{0}\rangle$ denotes the state where all qubits are prepared in their ground state $\ket{0}$. Let $ \langle \Psi_0| \hat{O}|\Psi_0\rangle=\langle \hat{O}\rangle$ be the expectation value we want to estimate within a variance of $\eta$.  The three steps of the CPS are:
\begin{enumerate}
    \item Obtain rough estimates of the mean values of every Pauli string $ \langle\Psi_0|\hat{P}_j| \Psi_0\rangle=\langle \hat{P}_j\rangle$, $j=[1,N]$ by performing $O\left(\log{\left(\frac{N}{\sqrt{\eta}}\right)}\right)$ 
    projective measurements similar to the QEE approach. This step estimates $s_j=\sign{(a_j\langle \hat{P}_j\rangle)}$ i.e. the sign information of the Pauli strings.
\item In the second step, $\tau_j|\langle \hat{P}_j\rangle|$, $\tau_j\equiv s_j \epsilon |a_j|$ is directly encoded in the phase of the single qubit memory using a modified phase-kickback algorithm~\cite{Knill2007} together with
QSP techniques. Here, $\epsilon$ is a tuning parameter that is used to circumvent the general "modulo $2\pi$" ambiguity of phase estimation, without a substantial scaling of the QSP circuit depth, which we will detail below. The encoding is done sequentially for all Pauli strings in the decomposition, resulting in a final phase of $\sim \langle \hat{O}\rangle$ followed by a projective measurement of the single qubit state.
\item 
The previous step is repeated a number of times with a varying parameter $\epsilon$ to obtain the final estimate of $\langle \hat{O}\rangle$. 
\end{enumerate}

At the end of these three steps, $ \langle  {O}\rangle$ can be estimated up to a variance of $\eta$ with an overall sample complexity of $T=O\left(N/\sqrt{\eta}(\frac{\log{(N/\sqrt{\eta})}}{\log{\log{(N/\sqrt{\eta})}}})\right)$.
We have counted this scaling as the number of state preparation circuits $V$ required in the method, which also includes the extra state preparation circuits that are part of the QSP step as detailed below. 

The second step, as defined above, is the main step of CPS, which we will now describe in detail. Let us define the state $|\Psi_j\rangle\equiv \hat{P}_j|\Psi_0\rangle$.
Following the arguments of Ref.~\cite{Wang2019}, we consider the unitary  $\hat{U}_{P_j}=\hat{V}\hat{\Pi}_0\hat{V}^{\dagger}\hat{P}_j$, where  $\hat{\Pi}_0=\hat{I}-2|\pmb{0}\rangle\langle\pmb{0}|$ is a multi-qubit reflection operator and $I$ is the identity operator. It is seen that the action of $\hat{\Pi}_0$ is to provide a $\pi$ phase only to the $|\pmb{0}\rangle$ state.  
The action of $\hat{U}_{P_j}$ is a rotation by a principal angle $\theta_j=\arccos{(|\langle\Psi_0|\Psi_j\rangle|)}=\arccos{(|\langle \hat{P}_j\rangle|)}$ in the subspace spanned by $|\Psi_0\rangle$ and $|\Psi_j\rangle$. Consequently, the
state $|\Psi_0\rangle$ can be written as an equal superposition of eigenstates $|\theta_j^{\pm}\rangle$  of $\hat{U}_{P_j}$ with eigenvalues $e^{\pm i\theta_j}$, respectively.  
If it is possible to project onto one of these eigenstates, the standard phase kickback method could be used to encode the phase $\theta_j$ into a single auxiliary qubit. A similar approach was considered in Ref.~\cite{Wang2019} to have a better estimation of each individual Pauli string. Such an approach, however, still suffers from the same accumulation of shot noise as the QEE method from the classical summation of the Pauli string estimates. In addition, efficient projection onto the eigenstates is only possible if the mean values are bounded away from zero.  

\begin{figure}[h!] 
    \centering 
    \includegraphics[width=6.5cm]{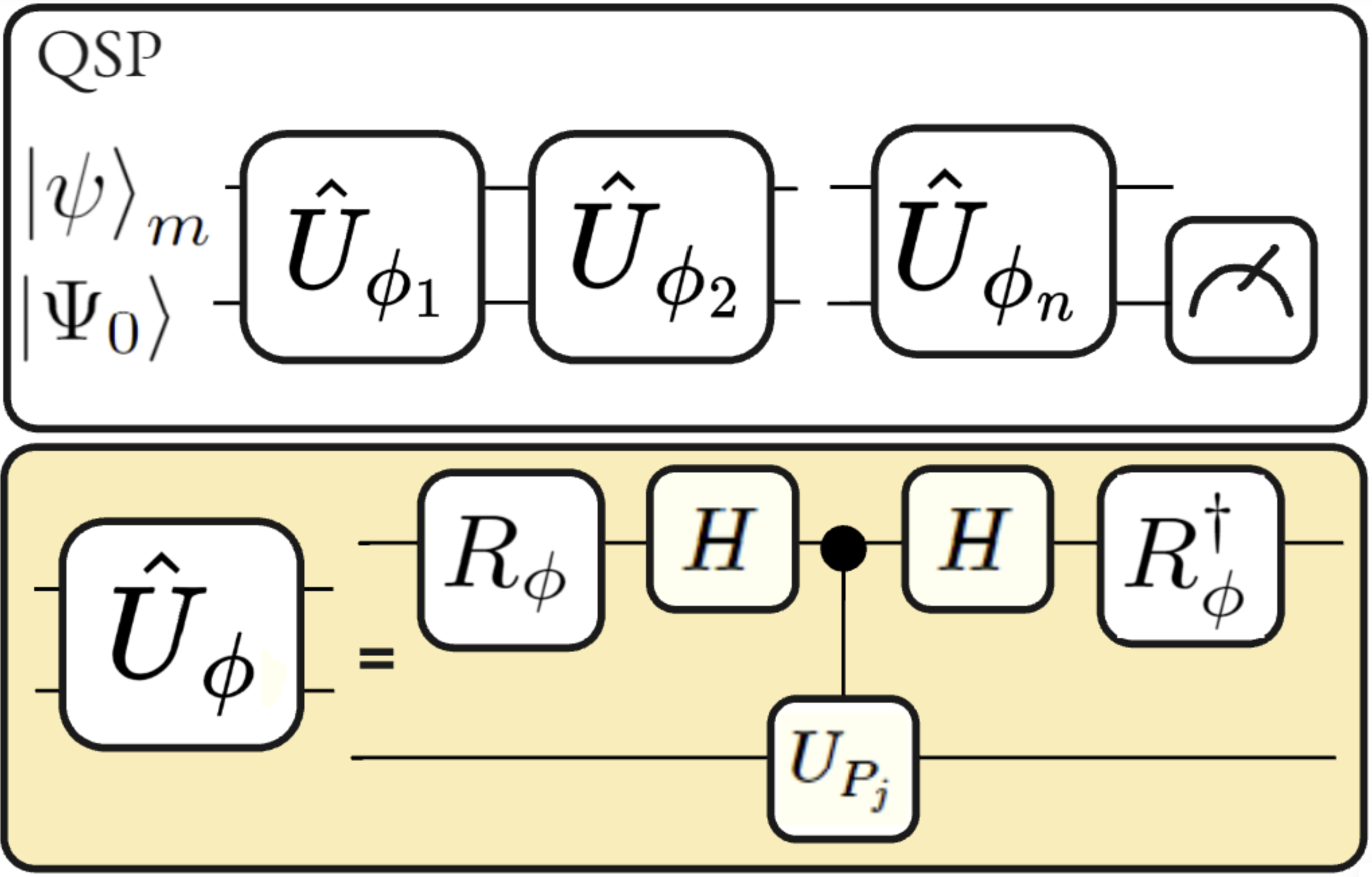}
    \caption{Quantum circuit realizing the QSP to encode one Pauli by a sequence of controlled unitaries $\hat{U}_{\phi_i}(\theta_j)$, $i=[1,n]$, built
    from Hadamard gates, $\hat{R}_{\phi}=e^{\frac{i\phi \hat{\sigma}_z}{2}}$,  and controlled $\hat{U}_{P_j}$. $\ket{\psi}_m=\alpha\ket{+}_m+\beta\ket{-}_m$ denotes the state of a single memory qubit. The state of the mutli-qubit device is reset after each encoding, resulting in modest requirements on the coherence time of the qubits. }
    \label{fig_qsp}
\end{figure}

The CPS method follows a different approach that allows for a direct encoding of the full observable $\hat{O}$ in order to circumvent the shot noise bottleneck. In addition, projection onto the eigenstates $|\theta_{j}^{\pm}\rangle$ is not required for the CPS method.  
To encode $\tau_j\cos({\theta}_j)$  (rather than $\theta_j$) in the phase of the memory qubit, we implement  a unitary $\hat{\mathcal{U}}_{P_j}$, which transforms $\hat{U}_{P_j}$ through QSP~\cite{PhysRevLett.118.010501}, such that $\hat{\mathcal{U}}_{P_j}\ket{\theta_j^{\pm}}=e^{i\tau_j\cos{( \theta_j)}}\ket{\theta_j^{\pm}}$. 
By iterating the basic building block depicted in Fig.~$\ref{fig_qsp}$ only $n=O(\log{(1/\epsilon_{QSP})}/\log{\log{(1/\epsilon_{QSP})}})$ times with the right choice of QSP phases $\phi_1\cdots\phi_n$, we compile a polynomial approximation of $\hat{\mathcal{U}}_{P_j}$ with $\tau_j=O(1)$, up to an error of $\epsilon_{QSP}$ (see Appendix~\ref{app_qsp}). The ability to adjust $\tau_j$ without incurring additional cost in terms of extra state preparations is essential for efficiently estimating an observable that scales with $N$ and for addressing the phase wrapping problem, which we discuss further below.
We note that the implementation of the controlled version of $\hat{U}_{P_j}$ (c$\hat{U}_{P_j}$) does not require controlled versions of the preparation circuit $\hat{V}$ but only controlled versions of the Pauli string $\hat{P}_j$ and $\hat{\Pi}_0$ operators. The control qubit for both operators will be the quantum memory qubit, $\ket{\psi}_m$, such that if $\ket{\psi}_m=\ket{0}_m$ no operation is applied on the target qubits, while if $\ket{\psi}_m=\ket{1}_m$, the operation is applied.

Applying the QSP sequence
on a (normalized) input state  $\left(\alpha\ket{+}_m+\beta\ket{-}_m\right)\otimes\ket{\Psi_0}$ gives the (unnormalized) state 
\begin{eqnarray}\label{2137}
&&\frac{1}{\sqrt{2}}\left({\alpha}|+\rangle_m+{\beta}e^{-2i\tau_j\cos{(\theta_j)}}|-\rangle_m\right)\nonumber \\
&&\otimes(\ket{\theta^+_j}+e^{-in\theta_j}\ket{\theta^-_j}) + \varepsilon_{QSP}(n)\ket{\xi},
\end{eqnarray}
where $\ket{\xi}$ is a general (normalized) error state, which can be an entangled state between the memory qubit and the qubits of the multi-qubit device. 

\begin{table*}[ht!]
\centering
\begin{tabular}{|c|c||  c|c|}
\hline 
Method&Number of  state preparations & Qubits &   Coherence time   \\ [0.5ex] 
 \hline
QEE &$O(N^2/\eta)$&Processing qubits &    $t_{prep}$   \\
 \hline
 QPEh & $O({N}^{3/2}/\sqrt{\eta})$&Processing qubits &  $t_{prep}O(\sqrt{N}/\sqrt{\eta}))$    \\
 \hline
CPS&$O\left(\frac{N}{\sqrt{\eta}}\frac{\log{(N/\sqrt{\eta})}}{\log{\log{(N/\sqrt{\eta})}}}\right)$ &Processing qubits &     $t_{prep}O\left(\frac{\log{(N/\sqrt{\eta})}}{\log{\log{(N/\sqrt{\eta})}}}\right)$  \\
&  
&Memory qubit     & $t_{prep}O\left(\frac{N}{\sqrt{\eta}}\frac{\log{(N/\sqrt{\eta})}}{\log{\log{(N/\sqrt{\eta})}}}\right)$    \\
\hline

\end{tabular}
\caption{ Comparison of resources for the QEE, QPE with the Heisenberg scaling per Pauli and CPS methods for a fixed target variance $\eta$.
We have quantified the necessary coherent time of the processing qubits of quantum device and the memory qubit in terms of the time required for a single state preparation $t_{prep}$, which is assumed to be the dominant scale. The observable is assumed to be decomposed into a summation of $N$ Pauli strings \eqref{eq:0}. We refer to the main text and Appendix~\ref{app_c} for further details)}
\label{tab_1}
\end{table*}

Repeating the above procedure for all Pauli strings in the decomposition of $O$ in a sequential manner, we prepare the single quantum memory (QM) qubit in a state 
\begin{eqnarray}
\label{1601}
&\approx&{\alpha}|+\rangle_m+{\beta}e^{-2i\sum_{j=1}^{N}\tau_j\cos{(\theta_j)}}|-\rangle_m\nonumber \\ 
&\approx& {\alpha}|+\rangle_m+{\beta}e^{-2i\epsilon\langle O \rangle}|-\rangle_m
\end{eqnarray} up to an error of $\sim N\varepsilon_{QSP}(n) $ assuming roughly the same approximation error for each Pauli string. Consequently, $\langle O \rangle$ is encoded in the phase of the QM qubit. 
\subsection{Phase wrapping}
Measuring the phase in Eq.~\eqref{1601} by a series of projective measurements would result in an estimation variance scaling as the standard quantum limit instead of the Heisenberg limit (see Appendix~\ref{app_qsp}). Moreover, since the phase encoding only provides an estimate of $\langle \hat{O} \rangle$ mod $2\pi$,  we are facing a phase wrapping problem. The observable we want to estimate can be written as $2\pi C+B$, where $C\in\mathbb{Z}$, $B\in \mathbb{R}$. So direct measurement will provide us with a precise estimate of $B$, losing the information about the amount of phase wraps $C$. To mitigate this issue and to achieve the Heisenberg scaling, we have included the factor of $\epsilon$ in the encoding. This brings us to the third and final step of the CPS.  
To control the phase wrapping, we use the \textit{sampling} approach introduced in 
Refs.~\cite{Higgins2009,Kimmel2015} to estimate $C$ and $B$ separately.
Instead of using fixed $\epsilon$ to encode $\theta_j$, we sample at multiple orders ${\epsilon}_l(\epsilon_0)\equiv 2^{l-1} \epsilon_0$, $l=1,2,\dots$  to gradually enclose on $\langle \hat{O} \rangle$.
This trick can be seen as sequentially estimating the digits of $\langle \hat{O}\rangle$. 
\par For estimating $C$, the parameter $\epsilon_0 \sim 1/N$ is chosen such that the condition $\sum_{j=1}^N \tau_j \cos{\theta_j} \leq 2\pi$ is satisfied. The QSP procedure with different ${\epsilon}_l(\epsilon_0)$ is repeated for $l \in [1, L]$, where $L = O(\log(1/\sqrt{\eta}))$. The estimation of $B$ is performed in a similar way, but with $\epsilon_0 = 1$. Note that naively scaling the variable down by a factor of $\epsilon_0 \sim 1/N$ and conducting standard phase estimation would require number of state preparations that scale with $N^2$. However, because we encode each Pauli string as a phase $e^{i\theta}$ with $\theta \in \{0,2\pi\}$ and QSP techniques allow us to realize maps between $\{0,2\pi\}\rightarrow\{0,2\pi\}$ with a nearly constant number of QSP iterations, we can estimate the observable $\hat{O}$ up to $O(1)$ variance using only $\log(N)$ state preparations for each Pauli string. Consequently, we find that with a total number of state preparations scaling as $O\left(\frac{N}{\sqrt{\eta}}\frac{\log{(N/\sqrt{\eta})}}{\log{\log{(N/\sqrt{\eta})}}}\right)$, the CPS method results in an estimate of $\hat{O}$ with a variance of $\eta$.



\section{Resources comparison}
In Tab.~\ref{tab_1}, we compare the required resources for the QEE, QPEh, and CPS methods: the required coherence time of the processing qubits and the number of state preparation circuits. Here QPEh referes to estimating each Pauli string with QPE and then sum the estimates classically to obtain $\langle O\rangle$~\cite{dutkiewicz2021,Kimmel2015,Wang2019}.
The coherence time is quantified by the time required for the state preparation circuit, $t_{prep}$.  
The scaling of the number of state preparations is the proper indicator for the complexity of the CPS method when the cost of implementing controlled reflection and Pauli operations does not scale faster than that of state preparation. This requirement is satisfied easily for the controlled Pauli operation when the $\hat{P}_j$ are $k$-local. On the other hand, the controlled reflection operation can be implemented using resources that scales linearly in the number of qubits on the hardware that supports native multi-qubit gates~\cite{molmer2011efficient,isenhower2011multibit,PhysRevLett.127.120501,PRXQuantum.1.020302}.

The QEE method only requires a coherence time of the multi-qubit device of $\sim t_{prep}$ since after each state preparation the qubits are measured. The QPEh method requires a coherence time scaling as $t_{prep}O(\sqrt{N}/\sqrt{\eta}))$ since each Pauli string is estimated using standard QPE. In comparison, the CPS method comes with different requirements on coherence time for the multi-qubit device and the single qubit memory. For the multi-qubit device, a modest increase in the coherence time compared to QEE which scales as $O\left(\frac{\log{(N/\sqrt{\eta})}}{\log{\log{(N/\sqrt{\eta})}}}\right)$ is required. This comes from the requirement of encoding each Pauli string using a logarithmically scaling number of QSP steps. On the other hand, the coherence time of the memory qubit has a similar scaling as QPEh up to a logarithmic factor. In return, the CPS provides a much better estimate of $\langle O \rangle$ than the QEE for a fixed number of state preparations. We find that 
\begin{eqnarray}\label{3004}
\frac{\sigma^2{(\hat{O}_{CPS})}}{\sigma^2{(\hat{O}_{QEE})}}\sim \frac{1}{T},
\end{eqnarray}
which shows that the CPS achieves a Heisenberg-like scaling of the variance in the number of state preparations compared to the standard quantum limit scaling of the QEE. CPS outperforms the QPEh too:
\begin{eqnarray}\label{3005}
\frac{\sigma^2{(\hat{O}_{CPS})}}{\sigma^2{(\hat{O}_{QPEh})}}\sim \frac{1}{N},
\end{eqnarray}
which is because CPS circumvents the accumulation of noise in the classical summation in QPEh. 

The limiting factor of both QEE and CPS will be the accumulation of operational errors in the final estimate of the observable. For both approaches, gate errors will reduce the accuracy of the final estimate. This reduction will have the same linear dependence on the number of Pauli strings in the observable decomposition for both approaches (see Appendix~\ref{ap_c}).
For the estimation of observable such as the energy of a $\text{H}_3^+$ molecule, which can be decomposed into $N= 59$ Pauli strings~\cite{crawford2021efficient} on $4$ qubits, an estimate of the energy with a variance of $\eta=1$ could be obtained for gate error probabilities  $\sim 4.5\cdot10^{-5}$ and at least $1.1\times 10^4$ state preparations using CPS. In comparison, QEE would require $\sim 3.48\times 10^3$ state preparations and QPEh $\sim 4.85\times 10^3$ to reach a similar variance. We do note, however, that the QEE approach would only require gate error probabilities on the order of $4.2\times 10^{-3}$, while QPEh $\ 1.8\times 10^{-4}$ (we refer to Appendix~\ref{app_d} for more details).  
To simulate larger molecules, the performance of current hardware needs to be improved for both our method and QEE. 
For example, the CPS simulation of a $\text{LiH}$ molecule with a decomposition of $\sim 630$ Pauli strings on $10$ qubits, requires $\sim  1.4\times 10^{5}$ state preparations and gate error probabilities $\sim 1.43 \times 10^{-6}$, while the QEE method requires $\sim 3.96\times 10^{5}$ state preparations and gate error probabilities $\sim 1.6\times 10^{-4}$ to achieve the same variance of $\eta=1$. The QPEh requires $1.69\times 10^4$ state preparations and gate error probabilities of a rate $\sim 2.1\times 10^{-6}$. Note, however, that the estimates for the required gate error probabilities are conservative since we assume that just a single error corrupts the estimation. For specific implementations and hardware models better bounds can likely be obtained.

\section{Summary}
In summary, we propose a new method (CPS)  to estimate the expectation values of multi-qubit observables.  The method uses the QSP technique to encode information from a multi-qubit processor into a single qubit quantum memory, which allows to overcome the shot noise bottleneck classically summing individually estimated Pauli strings. Compared to the QEE, the CPS obtains an Heisenberg limited scaling of the estimation variance  with the number of state preparations. This scaling represents an improvement of $~1/T$. In addition, CPS achieves an improvement scaling as $\sqrt{N}$ compared to estimating each Pauli string with standard QPE followed by classical summation. We note that this improvement has been estimated assuming that there is on the order of $N$ non-commuting Pauli strings in the observable decomposition. If there are commuting sets of Pauli strings they can in principle be measured in parallel using the QEE method while the CPS method does not straightforwardly support parallel encoding of commuting Pauli strings. Thus, we imagine that potential trade-offs between commuting sets and non-commuting set of Pauli strings can be made, resulting in optimal strategies consisting of both methods depending on the specific observable. CPS can be considered as a new variation of QPE for complex observables where direct encoding of the observable as a unitary is not possible.

While the CPS is designed for the estimation of a general observable, we believe that it will, in particular, be relevant for  algorithms such as the variational quantum eigensolver where the observable is the energy of the system. In particular, for estimating molecular energies where the mapping from a fermionic system to a qubit system often introduces highly non-local terms and large Pauli string representations. We believe that platforms with native multi-qubit controlled gates such as Rydberg atoms~\cite{PhysRevLett.127.120501,PRXQuantum.1.020302,pelegri2022high} or systems with coupling to a common bus mode such as a cavity~\cite{PhysRevLett.114.110502,doi:10.1126/science.add9771} 
would be particularly suited due to their potential for implementing the controlled unitaries required for the CPS. 

Finally, we note that another algorithm, which also uses QSP techniques to tackle the shot-noise bottleneck was recently proposed in Ref.~\cite{PhysRevLett.129.240501}. Here $\tilde{O}(\sqrt{N/\eta})$ (where $\tilde{O}(\cdot) $ is used to hide the logarithmic factors) state preparations are needed to obtain an estimate within a variance of $\eta$ but $O(N\log{(N/\sqrt{\eta})}+N)$ ancillary qubits and a circuit depth of $\tilde{O}(1/\sqrt{\eta})$ are required. Reducing the amount of ancillary qubits
to $O(N+\log{(1/\sqrt{\eta})})$, the same variance can be achieved with $\tilde{O}(N/\sqrt{\eta})$ state preparation queries, which is similar to the CPS method. However, the CPS method  requires only one auxiliary memory qubit.

\begin{acknowledgments}
L.M. was  supported by the Netherlands Organisation for Scientific Research (NWO/OCW), as part of the Quantum Software Consortium program (project number 024.003.037 / 3368).
This work has received support from the European Union’s Horizon Europe research and innovation programme through the ERC StG FINE-TEA-SQUAD (Grant No. 101040729). This work is supported by the Dutch National Growth Fund (NGF), as part of the Quantum Delta NL programme.  J.B. acknowledge funding from the NWO Gravitation Program Quantum Software Consortium. J.B. acknowledges support from The AWS Quantum Discovery Fund at the Harvard Quantum Initiative


Funded by the European Union. Views and opinions expressed are however, those of the author(s) only and do not necessarily reflect those of the European Union or the European Commission. Neither the European Union nor the granting authority can be held responsible for them.
\end{acknowledgments}

\bibliographystyle{plainnat}
\bibliography{Bibliography}

\onecolumn
\appendix

\section{CPS method}\label{app_qsp}
\par In this section, we first provide an overview  of the basic notion of Quantum Signal Processing (QSP) \cite{PhysRevLett.118.010501} for the introduction of the Coherent Pauli Summation (CPS) method. 
\par  Let us consider a single qubit rotation of the form
\begin{eqnarray}\label{9012}
\hat{R}_{\phi}(\theta)=e^{-i\frac{\theta}{2}(\hat{\sigma}_x\cos{\phi}+\hat{\sigma}_y\sin{\phi})}
\end{eqnarray}
with an  angle $\theta$. This rotation can be considered as a computational module that computes a unitary function that depends on the selected parameter $\theta$, the input state and the measurement basis.
A sequence of such $n$ single qubit rotations can be expressed as
\begin{eqnarray}\label{9014}\hat{R}_{\phi_n}(\theta)\dots \hat{R}_{\phi_2}(\theta)\hat{R}_{\phi_1}(\theta)=
A(\theta)I+i B(\theta)\hat{\sigma}_z+i C(\theta) \hat{\sigma}_x+i D(\theta) \hat{\sigma}_y. \end{eqnarray}
With a specific choice of different parameters $\Vec{\phi}$, it is possible to compute  more general  functions of $\theta$  in terms of $A(\theta)$, $B(\theta)$, $C(\theta)$, and $D(\theta)$ being  polynomials of, at most, degree $n$. Often it is enough to use a partial set of $(A,B,C,D)$, for example $(A,C)$ (see Ref.~\cite{PhysRevX.6.041067} for more details). For this operation, the following theorem holds
\begin{theorem}\cite{PhysRevLett.118.010501}
For any even $n>0$, a choice of real functions $A(\theta)$, $C(\theta)$ can be implemented by some $\Vec{\phi}\in R^n$ if and only if all these are true:
\begin{itemize}
    \item For any $\theta\in R$, 
    \begin{eqnarray}\label{1522}
    A^2(\theta)+C^2(\theta)\geq 1,\quad \text{and}\quad A(0)=1,
    \end{eqnarray}
    \item $A(\theta)=\sum\limits_{k=0}^{n/2}a_k\cos{(k\theta)}$, $\{a_k\}\in R^{n/2+1}$, $C(\theta)=\sum\limits_{k=0}^{n/2}c_k\cos{(k\theta)}$, $\{c_k\}\in R^{n/2}$.
\end{itemize}
Moreover, $\Vec{\phi}$ can be efficiently computed from $A(\theta)$, $C(\theta)$.
\end{theorem}

Furthermore, given a unitary $\hat{U}$ with eigenstates 
$\hat{U}|\Psi_0\rangle=\frac{1}{\sqrt{2}}\sum_{\pm}e^{\pm i\theta}|\theta_{\pm}\rangle$, 
 a quantum circuit $\hat{\mathcal{U}}=\sum_{\pm}e^{ih(\pm\theta)}|\theta_{\pm}\rangle\langle \theta_{\pm}|$
can be constructed, where $h(\theta)$ is a real function.
To this end we need a $\vec{\phi}$ dependent  version of $\hat{U}$, namely
\begin{eqnarray}
    &&\hat{U}_{\phi}=(e^{-i\frac{\phi}{2}\hat{\sigma}_z}\otimes \hat{1})\hat{U}_0(e^{i\frac{\phi}{2}\hat{\sigma}_z}\otimes \hat{1}),\\\nonumber
    &&\hat{U}_0=\sum\limits_{\pm}e^{\pm i\theta/2}\hat{R}_0(\pm\theta)\otimes\ket{\theta_{\pm}}\bra{\theta_{\pm}}.
\end{eqnarray}
An operator $\mathcal{\hat{V}}=\hat{U}_{\phi_n}\hat{U}_{\phi_{n-1}}\dots \hat{U}_{\phi_1}$ approximating $\hat{\mathcal{U}}$ is introduced. Applying it to the input state $\ket{+}\ket{\theta_{\pm}}$ and post selecting on measuring $\bra{+}$, we get $\bra{+}\mathcal{\hat{V}}\ket{+}\ket{\theta_{\pm}}=(A(\theta_{\pm})+iC(\theta_{\pm}))\ket{\theta_{\pm}}$ with  the worst case success probability
$ p=\min\limits_{\theta}{|\langle +|\mathcal{\hat{V}}|+\rangle|^2}=\min\limits_{\theta}{|A(\theta_{\pm})+iC(\theta_{\pm})|^2}$.
The second theorem holds:
\begin{theorem}\label{thm_2}(quantum signal processing)\cite{PhysRevLett.118.010501}
Any real odd periodic function $h:(-\pi,\pi]\longrightarrow(-\pi,\pi]$ and even $n>0$, let
$A(\theta), C(\theta)$ be real Fourier series in $\cos{(k\theta)}$, $\sin{(k\theta)}$, $k=0,\dots, n/2$, that approximate 
\begin{eqnarray}\max\limits_{\theta\in R}{|A(\theta)+i C(\theta)-e^{ih(\theta)}|}\leq \varepsilon_{QSP}.
\end{eqnarray}
Given $A(\theta), C(\theta)$, one can efficiently compute $\vec{\phi}$ such that $\langle +|\mathcal{\hat{V}}|+\rangle$ applies $\hat{U}_{\phi}$ a number $n$ times to approximate $\hat{\mathcal{U}}$ with success probability $p\geq 1-16 \varepsilon_{QSP}$ and the distance $\max\limits_{\ket{\Psi}}{\|(\langle +|\mathcal{\hat{V}}|+\rangle-\hat{\mathcal{U}})\ket{\Psi_0}\|}\leq 8 \varepsilon_{QSP}$.
\end{theorem}
At this point, we note that we are not conditioning on a projection onto the $\ket{+}$ state in the CPS method. Consequently, the notion of a success probability $p$ is not really valid in the CPS method and $p$ instead turns into an error of the approximation of the unitary by the QSP method, as we will show below. For now, we, however, keep the notion of success probability to relate to existing literature.   

The query complexity of the
methodology is exactly the degree $n$ of optimal trigonometric polynomial approximations to $e^{ih(\theta)}$ with error $\varepsilon_{QSP}$.
It is mentioned, that $A(\theta)$, $C(\theta)$ satisfying the second theorem, in general, do not satisfy \eqref{1522}. Thus the rescaling is provided
\begin{eqnarray}\label{1523}
    &&A_1(\theta)=A(\theta)/(1+\varepsilon_{QSP}),\quad C_1(\theta)=C(\theta)/(1+\varepsilon_{QSP}),\\\nonumber
    &&|A_1(\theta)+i C_1(\theta)-e^{ih(\theta)}|\leq \varepsilon_{QSP}/(1+\varepsilon_{QSP})+\varepsilon_{QSP}<2\varepsilon_{QSP}.
    \end{eqnarray}
Hence, the success probability of the method is at least $1-2\varepsilon_{QSP}$.
\par In this paper, we are using the QSP technique to approximate $\exp{(i\sum_{j=1}^{N}\tau_j \cos{[\arccos{(|\langle P_j\rangle|)}]})}$, where we denote $\tau_j=s_j \epsilon |a_j|$, $s_j=\sign({a_j\langle P_j\rangle)}$ and $a_j$ are the coefficients of \eqref{eq:0}.
To this end, we select $h(\theta_j)=\tau_j \cos{(\pm \theta_j)}$, where   the input is $\theta_j=\arccos{(|\langle P_j\rangle|)}$.
Then for every $\theta_j$, we can write
\begin{eqnarray}\label{1528}
&&A(\theta)=\cos{(\tau\cos{(\theta)})},\quad C(\theta)=\sin{(\tau\cos{(\theta)})}.
\end{eqnarray}
We can use the real-valued Jacobi-Anger expansion \cite{abramowitz1965formulas} to rewrite the later functions as series:
\begin{eqnarray}\label{1641}
&&  \cos(\tau \cos{(\theta)}) \equiv J_0(\tau)+2 \sum_{m=1}^{\infty}(-1)^m J_{2m}(\tau) \cos(2m \theta),\\\nonumber
&&  \sin(\tau\cos{(\theta)}) \equiv -2 \sum_{m=1}^{\infty}(-1)^m J_{2m-1}(\tau) \cos\left[\left(2m-1\right)\theta\right],
\end{eqnarray}
where $J_{m}(\tau)$ is the $m$-th Bessel function of the first kind. The Bessel function $J_{m}(\tau)$  is bounded, for real $\tau$ and integer $m$ as
\begin{eqnarray}\label{1642}
&&  |J_{m}(\tau)|\leq \frac{1}{|m|!}\Big|\frac{\tau}{2}\Big|^{|m|}.
\end{eqnarray}
For $m\geq k$ the factorial $m!$ grows faster then the exponential $|\tau|^m$. As the latter decays with $m$,  good approximations are obtained \eqref{1641} at $m > n/2$.
 The sum from $m = k$ can be bounded by the first term:
\begin{eqnarray}
\sum_{m = k}^{\infty} \frac{1}{m!} \left( \frac{|\tau|}{2} \right)^m < \frac{1}{k!} \left( \frac{|\tau|}{2} \right)^{k} \cdot \sum_{m = 0}^{\infty} \left( \frac{|\tau|}{2k} \right)^m< \frac{1}{k!} \left( \frac{|\tau|}{2} \right)^{k},
\end{eqnarray}
where we used that since 
\begin{eqnarray}\label{1536}
    |\tau| \leq  n_{QSP}/2=k,
    \end{eqnarray}
    the ratio is
$|\tau|/2k < k/2k < 1$,
so the geometric series converges and is less than one.
Then the upper bound is~\cite{berry2015hamiltonian}:
\begin{eqnarray}\label{1642_0}
&&\varepsilon_{QSP}\leq 2 \sum_{m=k}^{\infty}|J_{m}(\tau)|\leq 4 \sum_{m=k}^{\infty}\frac{1}{m!}\Big|\frac{\tau}{2}\Big|^{m}<\frac{4}{k!}\Big|\frac{\tau}{2}\Big|^{k}<
4\left(\frac{e|\tau|}{2k}\right)^{k}.
\end{eqnarray}
The last inequity is due to the Stirling's approximation $l!>(l/e)^l$, for $l\geq 1$.
Then we can deduce
\begin{eqnarray}\label{931}
    k=O\left(\frac{\ln{(1/\varepsilon_{QSP})}}{\ln{k}}\right).
\end{eqnarray}
The logarithm in the denominator complicates direct solving. However, this suggests an iterative relation, that is a characteristic of the Lambert $W(z)$ function:
\begin{eqnarray}\label{932}
    W(z)e^{W(z)}=z.
\end{eqnarray}
We attempt to rewrite our expression into the latter form. Let us introduce the notation $E=1/\varepsilon_{QSP}$. Then, we can rewrite \eqref{931} as $k\ln{k}=O(E)$.
Changing the variables $y=\ln{k}$, $k=e^y$, we get $ ye^y=O(E)$. This is similar to the defining equation \eqref{932} of the Lambert $W(E)=y$ function. Then $k\approx e^{W(E)}$, holds. For large values of $E$, the Lambert $W(z)$ function can be approximated as:
\begin{equation}
W(E) = \ln E - \ln \ln E + \mathcal{O}(1).
\label{eq:W_approx}
\end{equation}
Then we can write
\begin{eqnarray} \label{1236}
k=e^{\ln E- \ln \ln E + \mathcal{O}(1)}=\frac{E}{\ln{E}}e^{O(1)}=O\left(\frac{E}{\ln{E}}\right)=O\left(\frac{\log{(1/\varepsilon_{QSP})}}{\log{\log{(1/\varepsilon_{QSP})}}}\right).
\end{eqnarray}
The error of the approximation of every
$\exp{(i\tau_j \cos{(\theta_j)})}$, $j=[1,N]$ is scaled super-exponentially 
\begin{eqnarray}\label{1643}
\varepsilon_{QSP}\leq O\left(\left(\frac{e\min|\tau_j|}{2k}\right)^{k}\right),
\end{eqnarray}
and the amount of quarries, taking into account \eqref{1536} and \eqref{1236}, is \cite{PhysRevLett.118.010501}:
\begin{eqnarray}\label{9011_0}
n_{QSP}=O\left(\max\limits_{j}{|\tau_j|}+\frac{\log{(1/\varepsilon_{QSP})}}{\log{\log{(1/\varepsilon_{QSP})}}}\right).
\end{eqnarray}
This is the number of times the QSP circuit must repeat to approximate the target function with an error $\varepsilon_{QSP}$.
We  demand  the standard deviation $\sqrt{\eta}$ of the estimate $\hat{O}$ to be greater than the QSP errors:
\begin{align}
    \varepsilon_{QSP}\leq \frac{\sqrt{\eta}}{N},
\end{align}
which require the number of steps in each QSP protocol to be 
\begin{eqnarray}\label{1417}
n_{QSP}=O\left(\max\limits_{j}{|\tau_j|}+\frac{\log{(\frac{N}{\sqrt{\eta}})}}{\log{\log{(\frac{N}{\sqrt{\eta}})}}}\right).
\end{eqnarray}
\par If we select $\ket{+}_m\otimes\ket{\Psi_0}$ as the initial state, the action of the ideal QSP unitary $\mathcal{U}^{{\rm ideal}}_{P_j}$ on this state is the following:
\begin{align}
\hat{\mathcal{U}}^{{\rm ideal}}_{P_j} \ket{+}_m\otimes\ket{\Psi_0} = \ket{+}_m \otimes \sum\limits_{\pm}e^{i\tau_j \cos{(\pm\theta_j)}}\ket{\theta_{\pm}}.
\end{align}
Note that $\ket{\Psi_0}$ is an equal superposition of two eigenstates $\ket{\theta_j^{\pm}}$ of $U_{P_j}$, with the eigenvalues $\pm\theta_j$, respectively. However, since $\mathcal{U}_{P_j}^{\rm ideal}$ has eigenvalues that are cosine transforms of the eigenvalues of $U_{P_j}$, the initial state $\ket{\Psi_0}$ is invariant under the action of  $\mathcal{U}_{P_j}^{\rm ideal}$. 
Moreover, according the Theorem \ref{thm_2}, choosing the initial state of the ancilla as $\ket{+}_m$, we select the functional transformation given by $A(\theta)+iC(\theta)$.
\par On the other hand, selecting the initial state of the ancillary qubit as $\ket{-}_m$, we obtain
\begin{align}
\mathcal{U}^{{\rm ideal}}_{P_j} \ket{-}_m\otimes\ket{\Psi_0} = \ket{-}_m \otimes \sum\limits_{\pm}e^{-i\tau_j \cos{(\pm\theta_j)}}\ket{\theta_{\pm}}.
\end{align}
In this case, the functional transformation we select is $A(\theta)-iC(\theta)$.
Hence, setting the initial state of the memory ancilla as 
\begin{align}
\ket{\psi}_m = \alpha \ket{+}_m + \beta\ket{-}_m,
\end{align}
we have the following action:
\begin{align}
    \mathcal{U}^{{\rm ideal}}_{P_j} \ket{\psi}_m\otimes\ket{\Psi_0} = \alpha\ket{+}_m \otimes \sum\limits_{\pm} e^{i\tau_j \cos{(\pm\theta_j)}}\ket{\theta_j^{\pm}} + \beta \ket{-}_m \otimes \sum\limits_{\pm}e^{-i\tau_j \cos{(\pm \theta_j)}}\ket{\theta_j^{\pm}}.
\end{align}
Repeating the QSP encoding for each Pauli string $P_j$, we encode the observable mean $\langle O\rangle$ in the phase of the memory qubit.
\par Since the QSP method is just approximating the function from the random variable, we have to take into account the error accumulation in the QSP approximation.
If we apply the QSP sequence
on a (normalized) input state  $\ket{\psi}_m\otimes\ket{\Psi_0}$ we get the (unnormalized) state 
\begin{eqnarray}\label{2137_1}
&\sim&\left(e^{i\tau_j\cos{(\theta_j)}}{\alpha}|+\rangle_m+{\beta}e^{-i\tau_j\cos{(\theta_j)}}|-\rangle_m\right)\otimes(e^{in\theta_j/2}\ket{\theta^+_j}+e^{-in\theta_j/2}\ket{\theta^-_j}) + \varepsilon_{QSP}(n)\ket{\xi},
\end{eqnarray}
where $\ket{\xi}$ is a general (normalised) error state, which can be an entangled state between the memory qubit and the qubits of the multi-qubit device. We reinitialize the computational qubit by tracing it out.  
After $N$ rounds of such QSP encoding, with probability $(1-\varepsilon_{QSP}(n))^N$, $\epsilon_{QSP}(n)<<1$  the state encoded in the memory qubit is:
\begin{eqnarray}
   |{\Phi}_N\rangle
   &\sim&
  {\alpha}e^{i\sum_{j=1}^{N}\tau_j\cos{(\theta_j)}}|+\rangle_m+{\beta}e^{-i\sum_{j=1}^{N}\tau_j\cos{(\theta_j)}}|-\rangle_m.
    \label{1254_5}
\end{eqnarray}
With probability $(1-(1-\varepsilon_{QSP}(n))^N) \sim N\varepsilon_{QSP}(n)$  we assume that don't encode any reasonable information i.e. that the memory qubit is depolarized due to erroneous QSP encoding of the Pauli strings.
\par Selecting $\alpha=\beta$, we can rewrite the latter state as follows:
\begin{equation}
  \ket{{\Phi}_N}\sim \cos{{\Phi}_N}
   |0\rangle_m+i\sin{{\Phi}_N}|1\rangle_m,
    \label{1058_6}
\end{equation}
where the phase ${\Phi}_N\equiv \sum_{j=1}^{N}\tau_j\cos{(\theta_j)}= \langle O\rangle$ is a target mean value. Hence, we need to estimate $\Phi_N$ which will provide $\langle O\rangle$. 
The probabilities to measure $|0\rangle$ and $|1\rangle$ are the following:
\begin{equation}
    P(X=0|{\Phi}_N)=\frac{1}{2}(1+\cos{(2{\Phi}_N)}),\quad P(X=1|{\Phi}_N)=\frac{1}{2}(1-\cos{(2{\Phi}_N)}),
    \label{1058_7}
\end{equation}
that gives a precise estimates of the modulus of the phase for a sufficient number of iterations. Measuring in the $y$-basis, we get the probabilities:
\begin{equation}
    P(Y=0|{\Phi}_N)=\frac{1}{2}(1-\sin{(2{\Phi}_N)}),\quad P(Y=1|{\Phi}_N)=\frac{1}{2}(1+\sin{(2{\Phi}_N)}).
    \label{1058_7_1}
\end{equation}

 Since we don't know the preparation circuit of the state $|{\Phi}_N\rangle$, we can't use QPE to estimate the phase directly. To estimate
$P(X=0|\Phi_N)$, $P(Y=0|\Phi_N)$ and then to estimate the total accumulated phase, we need to repeat all the procedure of encoding $\ket{\Phi_N}$ in the QM $M_q$ times, every time doing the projective measurement on $\ket{\Phi_N}$. Using the probability estimates, we can obtain the estimates of \textit{sine} and \textit{cosine} functions of ${\Phi}_N$.
\par  Using these estimates, we can  estimate $\Phi_N$. This can be done, for example, by introducing the notation 
\begin{equation}
    {Q}\equiv \frac{1-2{P}(Y=0| {\Phi_N})}{2{P}(X=0| {\Phi_N})-1}={\tan}{(2{\Phi_N})}.
    \label{21}
\end{equation}
The estimate of ${\Phi}_N$ is denoted by $\hat{\Phi}_N$. Then the variance of $\hat{\Phi}_N$ is the following:
\begin{equation}
\sigma^2\hat{\Phi}_N\approx\left(\frac{\partial {\Phi_N}}{\partial Q}\right)^2\sigma^2\hat{Q}=\left(\frac{1}{2(1+Q^2)}\right)^2\sigma^2\hat{Q}.
    \label{22_1}
\end{equation}
The variance of $\hat{Q}$ can be written as follows
\begin{equation}
    \sigma^2\hat{Q}\approx\sigma^2\hat{P}(X=0| {\Phi}_N)\left(\frac{2(1-2\hat{P}(Y=0| {\Phi}_N))}{(2\hat{P}(X=0| {\Phi}_N)-1)^2}\right)^2+\sigma^2\hat{P}(Y=0| {\Phi}_N)\left(\frac{2}{2\hat{P}(X=0| {\Phi}_N)-1}\right)^2.
    \label{22}
\end{equation}
Since we have a Bernoulli distributed random variables, the variances are
\begin{equation}
    \sigma^2\hat{P}(X=1| {\Phi}_N)=\frac{\hat{P}(X=0| {\Phi}_N)\hat{P}(X=1|{\Phi}_N)}{M_q},\quad \sigma^2\hat{P}(Y=1| {\Phi}_N)=\frac{\hat{P}(Y=0| {\Phi}_N)\hat{P}(Y=1| {\Phi}_N)}{M_q},
    \label{24_2}
\end{equation}
where $M_q$ is the amount of repreparations of the state $|{\Phi_N}\rangle$.
Finally, the variance \eqref{22_1} can be written as follows:
\begin{equation}
    \sigma^2(\hat{\Phi}_N)\approx\frac{3 + \cos{(8{\Phi}_N)} }{16 M_q}\leq \frac{1}{4 M_q}.
    \label{22_3}
\end{equation}
Note that this this variance is calculated for a fixed $\epsilon$ parameter. Then the variance of the estimate of the target observable is
\begin{eqnarray}
 \sigma^2(\langle \hat{O}\rangle)= \sigma^2(\hat{\Phi}_N) \epsilon^{-2} \sim \frac{1}{ \epsilon^2 M_q}.
\end{eqnarray}
One can see that estimating the observable in this way we only achieve the standard quantum limit. Moreover, there is a problem of phase wrapping since we can only resolve phases mod $2\pi$. To resolve these issues, we adopt the technique of robust phase estimation from Refs.~\cite{Higgins2009,Kimmel2015} as detailed below.

\section{Phase Wrapping Control and Heisenberg Scaling}\label{app}
\par 
The state  encoded in the memory qubit contains the phase 
\begin{eqnarray}\label{1149}
    \Phi=2\pi C+B,\quad C\in \mathrm{N}, \quad B\in \mathrm{R},
\end{eqnarray}
accumulated over $N$ rounds of encoding. In our method, since $\langle O\rangle$ is directly encoded as the rotation angle of a single qubit, expectation values differing in 
$C$  would be equivalent. An initial attempt to resolve this difficulty is to artificially scale 
$\langle O\rangle$ during the QSP step and rescale the estimate at the end of the protocol, using the phase estimation methods proposed by Higgins et al.~\cite{Higgins2009} and later generalized by Kimmel et al.~\cite{Kimmel2015}.

We introduce the notation 
\begin{eqnarray}\label{9011}
\Phi_l(\epsilon_0)\equiv \sum_{j=1}^{N}\tau^{(l)}_j(\epsilon_0)\cos{(\theta_j)},
\end{eqnarray}
  where we denote $\tau^{(l)}_j(\epsilon_0)=2^{l-1}{\epsilon}_0 s_j|a_j|$ such that $\Phi_0({\epsilon}_0)\leq 2\pi$ holds.   In ~\cite{Kimmel2015} the estimates  $\hat{\Phi}_l(\epsilon_0)$, $l=[1, d]$ of \eqref{9011}  are  obtained from $M_l=\alpha+\gamma(d-l)$, $\gamma>2$, $\alpha>0$ repetitions of the measurement circuits described in the previous section. 
The probability of making an error on every step is
\begin{eqnarray}
  && p_e(\Phi_{l}(\epsilon_0))\equiv P\Big[\hat{\Phi}_{l}(\epsilon_0)-\Phi_{l}(\epsilon_0)\geq \frac{\pi}{2}\vee \hat{\Phi}_{l}(\epsilon_0)-\Phi_{l}(\epsilon_0)<- \frac{\pi}{2}\Big],
\end{eqnarray}
and it is upper bounded by:
 \begin{eqnarray}
  && p_e(\Phi_{l}(\epsilon_0))<\frac{1}{\sqrt{2\pi M_l}2^{M_l}}.
\end{eqnarray}
On the last step $d=[\log_2{1/\sqrt{\eta}}]$ one obtains $\widehat{\epsilon_0\langle {O}\rangle}$ as an estimate of $\epsilon_0\langle O\rangle$ with a target variance $\eta$. For the total amount of state preparations $T$ the  variance  is $\sigma^2(\widehat{\epsilon_0\langle {O}\rangle})=\eta\sim O(T^{-2})$.
\par In contrast to the original method, where the initial encoded phase was assumed to be less than $2\pi$, we introduce a parameter $\epsilon_0$ that is sufficiently small to confine the phase within the desired range.  However, this modification prevents direct application of the original algorithm, as the variance of the estimate $\langle \hat{O}\rangle$ would then scale as $O(\epsilon_0^{-2} T^{-2})=O(N^2/T^2)$, preventing the attainment of  Heisenberg scaling. To overcome this difficulty, we implement the described method in two stages, estimating $C$ and $B$ in \eqref{1149} separately as it is shown below. Importantly, since QSP techniques allow us to encode each  Pauli string as a phases between $\{0,2\pi\}$ with a nearly constant number of QSP iterations, we can estimate the observable $\hat{O}$ up to $O(1)$ variance using only $\log(N)$ state preparations for each Pauli string.
\par 
As we discussed in the introduction, we demand the overall QSP approximation error smaller than the target variance variance $\eta$ (i.e., $\varepsilon_{QSP}\leq \sqrt{\eta}/N$), so the number of steps $n_{QSP}$ in each QSP protocol is approximately constant with $N$  (see Eq. \eqref{1417}).
However, encoding each $\Phi_l({\epsilon}_0)$  requires a different number of QSP steps to achieve a target error of 
$\varepsilon_{QSP}$.  From condition \eqref{1536}, we derive the following constraint:
\begin{eqnarray}\label{2213}
  n^{(l)}_{QSP}/2\geq |\tau_l(\epsilon_0)|\equiv 2^{l-1} \epsilon_0  \max\limits_{j=[1,N]}{|a_j|}, \quad l\in \mathrm{N}.
\end{eqnarray}
This is the minimal amount of the QSP steps needed to guarantee \eqref{1643} and \eqref{9011_0} to hold. 
Thus, for each $\Phi_l({\epsilon}_0)$, the number of required QSP steps scales with $N$ as follows:
\begin{eqnarray}\label{1010}
   n^{(l)}_{QSP}=O\left(|\tau_l(\epsilon_0)|+\frac{\log{(\frac{N}{\sqrt{\eta}})}}{\log{\log{(\frac{N}{\sqrt{\eta}})}}}\right).  
\end{eqnarray}
\par 
In the first step of our procedure, we aim to estimate $C$, which represents the number of phase wraps in \eqref{1149}, with a target variance $\eta_c$. We assume that we wish to obtain an estimate of $\langle O \rangle$ with a variance $\eta\ll1$ which means that $\eta_c\ll1$. To estimate $C$, we perform $L=[\log_2{1/\sqrt{{\eta_c}}}]$ steps of the algorithm above and estimate $C$ as:
\begin{eqnarray}
    \hat{C}=\Bigg\lfloor \frac{\hat{\Phi}_L(\epsilon_0)}{2\pi} \Bigg\rfloor.\end{eqnarray}
  If no errors occur in the principal range estimates for all 
  $\Phi_{l}(\epsilon_0)$, $l\in[1,L]$, then the maximum error in the estimate of $C$ is $2\pi/2^L$. 
We bound the variance of the estimate $\hat{C}$ of $C$ as:
\begin{eqnarray}
    \sigma^2(\hat{C})\leq (1-p_{max}(M_L))\left(\frac{2\pi}{2^L}\right)^2+\sum\limits_{j=1}^L \left(\frac{2\pi}{2^j}\right)^2p_{max}(M_j).
\end{eqnarray}
Doing the calculations, we get
\begin{eqnarray}
    \sigma^2(\hat{C})\leq \frac{\pi^2}{4^L}\left(4+\frac{1}{\sqrt{2\pi\alpha} 2^{\alpha}}\frac{16}{2^{\gamma}-2^2}\right).
\end{eqnarray}
Note that the condition $\gamma>2$  prevents the sum growing faster then $4^{-L}$. 
The total number of estimates required for this stage is
\begin{eqnarray}\label{1042}
    T_c=2N \sum\limits_{l=1}^L n^{(l)}_{QSP} M_l\leq 2N n^{(1)}_{QSP}\sum\limits_{l=1}^L2^l M_j 
\leq 2^{L+1}N n^{(1)}_{QSP} (\alpha+\gamma).
\end{eqnarray}
Finally, the variance of the estimate $\hat{C}$ is
\begin{eqnarray}
    \sigma^2(\hat{C})\leq \frac{N^2(n^{(1)}_{QSP})^2}{T_c^2}4\pi^2(\alpha+\gamma) ^2\left(4+\frac{1}{\sqrt{2\pi\alpha} 2^{\alpha}}\frac{16}{2^{\gamma}-2^2}\right).
\end{eqnarray}
This approach ensures that the variance of the estimate is independent of  $\epsilon_0$.
\par Now, we can apply the above algorithm to estimate $B\in(-\pi,\pi)$ with the target variance of $\eta_b\ll 1$. We directly encode $\Phi_l(1)=2^lP_{sum}$ by choosing $\epsilon_0=1$. The amount of QSP steps $\tilde{n}^{(l)}_{QSP}$ is given by \eqref{1010} where $\epsilon_0=1$ holds. We perform $K=[\log_2{1/\sqrt{{\eta_b}}}]$ steps of the algorithm, and the estimate of $B$ is  given by:
\begin{eqnarray}
 \hat{B}=\hat{\Phi}_K(1)\mod 2\pi.
\end{eqnarray}
The difference to the previous step is that even if all the steps of the algorithms provide no error, our final estimate can still differ from $B$ by at most $2\pi/2^{K+1}$. Then
we bound the variance of the estimate $\hat{B}$ of $B$ as:
\begin{eqnarray}
    \sigma^2(\hat{B})\leq (1-p_{max}(M_K))\left(\frac{2\pi}{2^{K+1}}\right)^2+\sum\limits_{l=1}^K \left(\frac{2\pi}{2^l}\right)^2p_{max}(M_l).
\end{eqnarray}
Similarly to the previous step, we get
\begin{eqnarray}
    \sigma^2(\hat{B})\leq \frac{\pi^2}{4^K}\left(1+\frac{1}{\sqrt{2\pi\alpha} 2^{\alpha}}\left(3+\frac{16}{2^{\gamma}-2^2}\right)\right).
\end{eqnarray}
The total amount of estimates required for this stage is
\begin{eqnarray}\label{1043}
    T_b=2N \sum\limits_{l=1}^K \tilde{n}^{(l)}_{QSP} M_l\leq 2N \tilde{n}^{(1)}_{QSP}\sum\limits_{l=1}^K 2^l M_l
\leq 2^{K+1}N \tilde{n}^{(1)}_{QSP} (\alpha+\gamma),
\end{eqnarray}
and the variance scales as
\begin{eqnarray}
    \sigma^2(\hat{B})\leq \frac{N^2(\tilde{n}^{(1)}_{QSP})^2}{T_b^2}4\pi^2(\alpha+\gamma) ^2\left(1+\frac{1}{\sqrt{2\pi\alpha} 2^{\alpha}}\left(3+\frac{16}{2^{\gamma}-2^2}\right)\right).
\end{eqnarray}
\par The total amount of state preparations needed for the CPS method is given by:
\begin{eqnarray}\label{1501}
   T&=&N  n_1 +T_c+T_b.
\end{eqnarray} 
Here, $N n_1$
represents the small number of projective measurements performed beforehand to roughly estimate the mean values of each Pauli string, where  $N  n_1 << T_c,T_b$. As we write in the main text, this procedure is needed for the Pauli sign estimation. 
Then, using \eqref{1042} and \eqref{1043}, we can write
\begin{eqnarray}
    T\approx 2N (\alpha+\gamma) (2^L{n}^{(1)}_{QSP} +2^K\tilde{n}^{(1)}_{QSP}).
\end{eqnarray}

The scaling of the QSP trials is the following
\begin{eqnarray}
    && {n}^{(1)}_{QSP}=O\left(|\tau_1(\epsilon_0)|+\frac{\log{(\frac{N}{\sqrt{\eta}})}}{\log{\log{(\frac{N}{\sqrt{\eta}})}}}\right)=O\left(\frac{1}{N}+\frac{\log{(\frac{N}{\sqrt{\eta}})}}{\log{\log{(\frac{N}{\sqrt{\eta}})}}}\right),\\\nonumber
    &&\tilde{n}^{(1)}_{QSP}=O\left(|\tau_1(1)|+\frac{\log{(\frac{N}{\sqrt{\eta}})}}{\log{\log{(\frac{N}{\sqrt{\eta}})}}}\right)=O\left(1+\frac{\log{(\frac{N}{\sqrt{\eta}})}}{\log{\log{(\frac{N}{\sqrt{\eta}})}}}\right).
\end{eqnarray}
Then substituting $L$ and $K$, we get
\begin{eqnarray}
    T\approx 2N (\alpha+\gamma) \left(\frac{1}{\sqrt{\eta_C}}{n}^{(1)}_{QSP} +\frac{1}{\sqrt{\eta_B}}\tilde{n}^{(1)}_{QSP}\right).
\end{eqnarray}
Since $C$ and $B$ are estimated independently, we can write
\begin{eqnarray}
    \sigma^2(\hat{\Phi})=4\pi^2 \sigma^2(\hat{C})+\sigma^2(\hat{B}).
\end{eqnarray}
The target variance is $\eta$, so we can require equal contribution of variances: $\eta_c=\eta/(8\pi^2)$, $\eta_b=\eta/2$. Then the variance of the target estimate is scaling as follows
\begin{eqnarray}
    \eta\approx\frac{8N^2}{T^2} (2\pi {n}^{(1)}_{QSP}+\tilde{n}^{(1)}_{QSP})^2(\alpha+\gamma)^2.
\end{eqnarray}
Hence, we can conclude that  the variance of the estimate for $\langle O\rangle$ is 
\begin{eqnarray}
    \sigma^2(\hat{O}_{CPS})=\frac{8(\alpha+\gamma)^2N^2 N^2_{QSP}}{T^2}=O\left(\frac{N^2 N^2_{QSP} }{T^2}\right),
\end{eqnarray}
where we used the notation $N_{QSP}\equiv 2\pi {n}^{(1)}_{QSP}+\tilde{n}^{(1)}_{QSP}=O\left(\frac{\log{(\frac{N}{\sqrt{\eta}})}}{\log{\log{(\frac{N}{\sqrt{\eta}})}}}\right)$. 
\par It is possible to include additive errors into analyses. For example, the state preparation and measurement errors are of this type. Then performing two families of experiments, the probabilities of measuring $X=0$, $Y=0$ are 
\begin{eqnarray}
   &&P(X=0|\epsilon_0{P}^{}_{sum},2^l)=\frac{1}{2}(1+\cos{(2^l \epsilon_0{P}^{}_{sum})})+\delta_x(2^l),\\\nonumber
&&    P(Y=0|\epsilon_0{P}^{}_{sum},2^l)=\frac{1}{2}(1-\sin{(2^l \epsilon_0{P}^{}_{sum})})+\delta_y(2^l),
    \label{1032}
\end{eqnarray}
respectively, where $\delta_x(2^l)$ and $\delta_y(2^l)$ are the additive errors. 
following  \cite{Kimmel2015} (Theorem 1.1.), if $\delta_l\equiv \sup\limits_{2^l}{\{|\delta_x(2^l)|,|\delta_y(2^l)|\}}<1/\sqrt{8}$ the same Heisenberg scaling holds.
\section[\appendixname~\thesection]{Comparison of CPS Method with Other Methods}\label{app_c}
\par 
To our best knowledge, the papers dedicated to the estimation of the mean of the complicated observable $O$ always use some variations of QEE or QPE methods~\cite{Kimmel2015,O_Brien2019,Wang2019}. That means, that every $ \langle P_i\rangle$ is estimated  individually, using one of these methods, and then summed classically.  As we mentioned in the text, if each  $\langle {P}_j\rangle$ is estimated with a variance  $\sigma^2(\langle\hat{P}_j\rangle)= \eta_p$,  the variance of the final estimate scales as  $\sigma^2(\hat{ O})\approx N\eta_p$, assuming roughly equal weights of the $\hat{P}_j$ in the decomposition of $\hat{O}$. That is why the only thing that can change from the method selection in the existing papers is the variance $\eta_p$.
\par In QEE every Pauli string is measured independently and then all the results are summed up. Hence, the variance of the estimate of $O$ is the following:

\begin{align}
    \sigma^2{(\hat{O}_{QEE})}&\equiv \sum\limits_{i=1}^{N}\frac{a_i^2\sigma^2{\hat{P}_i}}{n_c}=\frac{N}{M_c}\sum\limits_{i=1}^{N}a_i^2\left(1-\langle\Psi_0| P_i|\Psi_0\rangle^2\right)\sim\frac{N^2}{M_c},
    \label{eq:1049}
\end{align}
where $M_c=N n_c$ is the amount of projective  measurements  done in QEE. In QEE the amount of state preparations is equal to amount of measurements $T=M_c$. 
 By comparing the variances of the estimated observable by both methods, we get
 \begin{eqnarray}\label{3005}
\frac{\sigma^2{(\hat{O}_{CPS})}}{\sigma^2{(\hat{O}_{QEE})}}\sim  \frac{N^2_{QSP}}{T}.
\end{eqnarray}
\par The QPE scaling can significantly vary from the method.  In~\cite{Higgins2009,Kimmel2015}
 the Heisenberg scaling is achieved, where the variance of one Pauli mean scales as $\sigma^2(\hat{P}_{QPE_h})\sim c^2/T_P^2$. Then  the total variance is $\sigma^2(\hat{O}_{QPE_h})\sim Nc^2/T_P^2=N^3c^2/T^2$. Here $T=NT_P$ denotes the total amount of  controlled-$U$ used to estimate one Pauli. CPS beats this method
\begin{eqnarray}
    \frac{\sigma^2{(\hat{O}_{CPS})}}{\sigma^2{(\hat{O}_{QPE_h})}}\sim  \frac{N^2_{QSP}}{N}.
\end{eqnarray}
The Heisenberg QPE requires  implementation of  the observable  as a unitary (in contrast to CPS) and  enough circuit depth of the device.
If one is able to implement the whole observable $O$, then the Heisenberg scaling for its estimate is achieved.
\par In~\cite{wang2019accelerated}  a special coefficient $\alpha\in[0,1]$ is introduced to weight both methods. When $\alpha=0$ the method is nothing else as QEE, when 
  $\alpha=1$ the method is QPE. When $\alpha\in(0,1)$ some mean Pauli values are estimated according to QEE, some to QPE with $\alpha$ weights. One can see that if one have enough coherence time the variance of the method can't be better then QPE's, even if multiplied on some $\alpha$ dependent coefficient. Also, the result is again obtained for one Pauli, so in the end the total variance is formed by direct summation of $N$ variances for each Pauli estimate. 
\par  Knowing the standard deviation or the variance allows an immediate estimate of the accuracy of the estimated value. The Chebyshev's inequality can be written
\begin{eqnarray}
 \mathbb{P}\!\left(\Big|\hat{O}-{O}\Big|\geqslant \nu\!\right)\!\! \leqslant \frac{ \sigma^2(\hat{O})}{ \nu^2},
 \end{eqnarray}
 where $\nu >0$. We use the Chebyshev's inequality, since it does not require any assumptions on the probability distribution of the observable to be on a bounded support, but only existence of the finite variance. 
One can deduce that the amount of state preparations that guarantee the probability on the left hand side to be $p_0\in(0,1)$ if one uses different methods: 
\begin{eqnarray}
    &&T_{CPS}\sim \frac{N N_{QSP}}{ \nu\sqrt{p_0}},\quad T_{QEE}\sim \frac{N^2}{ \nu^2p_0},\quad T_{QPE_h}\sim \frac{N^{3/2}}{ \nu\sqrt{p_0}}.
\end{eqnarray}
The CPS outperforms all methods without paying a price of possibility to implement $O$ as a unitary that in general is not realistic for big $N$.

\section{Cost of implementing CPS}\label{ap_c}
\par If we want the CPS method for observable estimation to be as generally applicable to the results of quantum algorithms, we need to satisfy the following two requirements. First requirement is that the controlled reflection and controlled Pauli operations required for $c\hat{U}_{P_j}$ are not more error-prone than the uncontrolled state preparation unitary $V$. Second, we need to take into account the effect of applying the state preparation unitaries sequentially for the CPS method, as opposed to a shot-noise limited multi-shot estimation scheme where the system is measured after each application of $V$. The difference between the sequential and multi-shot estimation protocols with respect to the accumulation of errors will be discussed in Appendix~\ref{app_d} below. In this section, we will focus on the implementation of controlled reflection and Pauli operations. 

Ideally, the implementation of the controlled operations should be as simple as possible, such that the state preparation step remains the main source of errors. The state preparation unitaries that are feasible on current platforms are of constant depth, and the circuit size is linear in the system's size $n$, leading to an error probability that scales with $O(n)$. At the first sight, implementing a controlled reflection with error probability scaling as $O(n)$ is problematic.
The conventional implementation of a controlled reflection operation without the use of any ancillary qubits requires $O(n^2)$ Toffoli and CNOT gates, while a $O(n)$ Toffoli implementation of the controlled reflection operation is possible with an additional $O(n)$ ancillary qubits \cite{nielsen2002quantum}. Hence, a conventional implementation of controlled reflection operator would be more resource-intensive than the state preparation unitary $V$, implying that the advantage of the CPS scheme comes at the cost of a significantly increased gate-count, compared to simply preparing the state of interest and measuring each Pauli separately. 

Luckily, the Rydberg atom array platform allows for the implementation of the controlled reflection operator using a number of gates that grows linearly in the system size \cite{molmer2011efficient,isenhower2011multibit}. The scheme relies on the three-level structure of neutral atoms to eliminate the need for the $O(n)$ ancillary qubits in the conventional implementation. Here, the long-lived hyperfine subspace encodes the qubit states, while  the highly-excited Rydberg state mediates strong and long-ranged dipolar interactions \cite{saffman2010quantum,saffman2016quantum}. 

As first proposed in Refs. \cite{molmer2011efficient,isenhower2011multibit}, a multiple control $C_n Z$ gate can be implemented by choosing different Rydberg states for the control and target qubits. This strategy has two favorable features. First, the control atoms, which are excited to the same Rydberg state, interact via shorter range van der Waals interactions and can therefore be trapped in close proximity. Second, the 
long-range nature of resonant dipolar interactions between Rydberg states with different symmetries allow any of the control atoms to blockade the dynamics of the target atom independently. Using such a configuration of the control and target registers, the implementation of the $C_n Z$ gate proceeds in an analogous way to the conventional CNOT gate implementation proposed in Ref. \cite{lukin2001dipole,jaksch2000fast}. First, each control atom \textit{not} in the $\ket{0}$ state is simultaneously transferred to the Rydberg state $\ket{R}$. Then, the target state goes through a $2\pi$ rotation within the associated $\ket{0}$ and $\ket{R}$ subspace. The $2\pi$ rotation is blockaded unless the control atoms are all in the corresponding $\ket{0}$ state. Finally, the control atoms are transferred back from the Rydberg state to the logical subspace. Thus, the state of the system acquires a $-\pi$ phase only if all control atoms are in the zero state. The error probability of the protocol depends on the state of the control register, which determines the number of control atoms that are transferred to the short-lived Rydberg state. Since on average $n/2$ control qubits are excited to the Rydberg state, the error probability of the gate grows approximately linearly with $n$.


\section{Gate Errors}\label{app_d}
In the previous deductions, we assumed perfect quantum gates. However, any quantum device will have non-negligible gate errors and it is therefore important to estimate the effect of these on the performance of the observable estimation. 

The cardinal task of the QSP method is to encode an estimate of the observable into the phase of the memory qubit. Therefore, we can simplify the discussion of gate errors by looking at the scenario of a faulty encoding of a phase into a single qubit. Let us assume that with probability $1-p$, $p\in[0,1]$ no error happens in the phase encoding and the true phase $\Phi_N$ is encoded in the single qubit. With the probability $p$ some gate errors happened, and the phase encoded in the memory qubit is an unpredictable value $\in[0,2\pi]$. 

The process is repeated many times in order to achieve an estimate of $\Phi_a$ similar to the description following Eq.~(\ref{1058_6}) in Appendix A above.  
However, instead of sampling from the true probability distribution corresponding to $\Phi_N$, we are instead sampling from the distributions
\begin{eqnarray}
    \label{1304}
    && P(X=0|{\Phi}_N)=\frac{(1-p)}{2}(1+\cos{(2{\Phi}_N)})+\frac{p}{2},\\\nonumber
   && P(Y=0|{\Phi}_N)=\frac{(1-p)}{2}(1-\sin{(2{\Phi}_N)})+\frac{p}{2},
\end{eqnarray}
where $\Phi_N$ is the phase encoded when no gate errors were acquired, while the second term corresponds to the case when a gate error happened, and we assume the memory qubit is left in a completely depolarised state.
We can follow the procedure of the previous section to introduce \eqref{21} with the updated probabilities from Eqs.~(\ref{1304}). All the derivations  are similar in this case, and we again arrive at a variance of the estimate, which scales as $ \sigma^2\hat{\Phi}_N\sim 1/M_q$. Note, however, that the estimated value with gate errors ($\hat{Q}_g$) will deviate from the true value without gate errors $\hat{Q}$ as
\begin{align}
    |\hat{Q}-\hat{Q}_g|\lesssim p.
\end{align}
in the limit of $p\ll1$. The effect of gate errors will thus lead to an inaccuracy in the estimate. This inaccuracy should be smaller than the targeted standard deviation ($\sqrt{\eta}$) of the estimation in order to be negligible.   
\par  We assume that the errors arising in our protocol are dominated by the errors emerging in the state preparation circuits. As we state in Appendix~\ref{ap_c}, we suppose that the number of gate operations in the state preparation circuit is roughly set by the number of qubits $n$. We then approximate the total probability of an error happening during the state preparation circuit as $\sim n p_1$, where $p_1\ll1$ is the average probability of making an error during a single gate operation (one-qubit or two-qubit gate).  
The probability of a gate error happening during one QSP round is then $p\sim 3p_1 n$, since it involves three state preparation circuits. From this, we can estimate that the required gate error probability should be low enough such that 
\begin{eqnarray}\label{1100}
  \eta\gtrsim 3\sqrt{2}N np_1 N_{QSP}
\end{eqnarray}
holds. 
\par  In the QEE method, we perform one state preparation per Pauli string. Following the arguments above, the total gate error probability is thus $p\sim n p_1$ while sampling on one Pauli string. The measurement of the Pauli string can either yield outcome $X=0,1$ and we can express the probability to get outcome $X=1$ as
\begin{eqnarray}
   && P(X=1)=(1-p) P(P_j=1)+p P(P_{error}=1),\\
\end{eqnarray}
where $P(P_j=1)$ is the probability from measuring on the correct state, while $P(P_{error}=1)$ is the probability from measuring on state with gate errors. 
The mean value of any Pauli string can be expressed as a  mathematical expectation of a Bernoulli variable. 
Then the variance is
\begin{eqnarray}
    \sigma^2 P_j= P(X=1)(1- P(X=1)),
\end{eqnarray}
 and the variance of a random variable $O$ estimated by $M_c$ rounds of the QEE method is
\begin{eqnarray}
    \sigma^2 O_{QEE}=\frac{N}{M_c}\sum\limits_{j=1}^N a_j^2(1-\langle P_j\rangle^2)=\frac{N}{M_c}\sum\limits_{j=1}^N a_j^2(1-P(X=1)^2).
\end{eqnarray}
As with the QCP method, gate errors will lead to an inaccuracy of the estimate on the order of $\sim Nnp_1$ assuming roughly equal weights of all Pauli strings in the decomposition. For this inaccuracy to be negligible for a target variance of $\eta$, we thus require that 
\begin{eqnarray}
    \sqrt{\eta}\gtrsim N n p_1.
\end{eqnarray}
\par 
Similarly to the CPS, we assume that for the QPEh  the probability of error is dominated by the errors of the state preparation circuit. The probability of a gate errors during  QPEh for a single Pauli estimation  is  $\sim 3n p_1$. Then we demand $
    \sqrt{\eta}\geq 3 N n p_1/\sqrt{\eta_1}$
    to hold, where $\eta_1$ is a variance of a single Pauli estimate. Let $\eta_1=\eta/N$, then we get
 \begin{eqnarray}
    \eta\geq 3 N^{3/2} n p_1.\end{eqnarray}
To estimate an observable, such as the energy of molecules like $\text{H}_3^+$, $\text{LiH}$, $\text{H}_2\text{O}$, and $\text{CH}_4$, which can be decomposed into $N = \{59, 630, 1085, 3887\}$ Pauli strings~\cite{crawford2021efficient} and require $n = \{4, 10, 12, 16\}$ qubits respectively, with a target variance of $\eta = 1$, the corresponding single-gate error probabilities are summarized in Table~\ref{tab:results1}. 
\begin{table}[h!]
\centering
\begin{tabular}{ccccc}
\toprule
\textbf{\( N\)} & \textbf{\( n \)} & \textbf{\( p_{1,\text{CPS}} \)} & \textbf{\( p_{1,\text{Heisenberg}} \)} & \textbf{\( p_{1,\text{QEE}} \)} \\
\midrule
59    & 4  & $4.49 \times 10^{-5}$ & $1.839\times 10^{-4}$  & $4.24\times 10^{-3}$     \\
630   & 10 & \( 1.43 \times 10^{-6} \) & \( 2.11 \times 10^{-6} \) & $1.59 \times 10^{-4}$ \\
1085  & 12 & \( 6.66 \times 10^{-7} \) & \( 7.77 \times 10^{-7} \) & $7.68 \times 10^{-5}$\\
3887  & 16 & \( 1.28 \times 10^{-7} \) & \( 8.50 \times 10^{-8} \)  & $1.61 \times 10^{-5}$ \\
\bottomrule
\end{tabular}
\caption{The average probability of making an error during a single gate operation $p_1$ to achieve a fixed variance for CPS, $QPE_h$ and QEE methods.}
\label{tab:results1}
\end{table}
The upper bounds of the amount of state preparations required for CPS method to the Heisenberg QPE given in Ref.~\cite{Kimmel2015} with $\alpha=1/2$, $\gamma=5/2$ constants that is stated to provide the best scaling and the QEE method are 
\begin{eqnarray}\label{1109}
   T_{CPS}=\frac{6 \sqrt{2} N (2\pi {n}^{(1)}_{QSP}+\tilde{n}^{(1)}_{QSP})}{ \sqrt{\eta}},\quad T_{QPE_h}= \frac{10.7 N^{3/2}}{ \sqrt{\eta}},\quad T_{QEE}= \frac{N^2}{ \sqrt{\eta}}.
\end{eqnarray}
The values of ${n}^{(1)}_{QSP}$ and $\tilde{n}^{(1)}_{QSP}$ are calculated numerically, solving \eqref{1642_0} for $k$ for different values of $\epsilon_0$. 
The numerical comparison of \eqref{1109} for the $\text{H}_3^+$, $LiH$, $H_2O$, $CH_4$ molecules is provided in Table~\ref{tab:results}.
\begin{table}[h!]
\centering
\begin{tabular}{lcccc}
\toprule
\textbf{Molecule} & \textbf{\( N \)} & \textbf{\( T_{{CPS}} \)} & \textbf{\( T_{QPE_h} \)} & \textbf{\( T_{QEE} \)} \\
\midrule
\( H_3^{+} \)   & 59    &  $1.11 \times 10^{4}$  &  $4.85 \times 10^{3} $ & $3.48 \times 10^{3} $    \\
\( LiH \)       & 630   & $1.40 \times 10^{5}$ & $1.69 \times 10^{5}$&  $3.97 \times 10^{5}$\\
\( H_2O \)      & 1085  & $2.50 \times 10^{5}$ &  $3.82 \times 10^{5}$ & \( 1.177 \times 10^6 \) \\
\( CH_4 \)      & 3887  & $9.73 \times 10^{5}$ & \( 2.593 \times 10^6 \) & \( 1.511 \times 10^7 \) \\
\bottomrule
\end{tabular}
\caption{The amount of state preparations $T_{CPS}$, $T_{QPE_h}$, $T_{QEE}$ for different molecules.}
\label{tab:results}
\end{table}
One can conclude that as expected QEE is the most tolerant to the gate errors, however paying a high price of the amount of state preparations for estimating the energies of the molecules with the bigger amount of Paulis. The CPS outperforms both methods in the amount of state preparations already on the $LiH$ molecule.

\end{document}